# NEW O-Cs FOR THE CLASSICAL CEPHEIDS RY CAS AND V LAC OVER A CENTURY (1873-2022)

**Guy Boistel**

GEOS (Groupe Européen d'Observation Stellaire), http://geos.upv.es/

**Abstract**

We give here new O-C diagrams of the two classical Cepheids RY Cas and V Lac, established on a large basis of times of maximum of light, which covers more than a century.

These two Cepheids present a variation of their period for which it is possible to calculate an annual rate. This rate equals +1.86 s/yr for RY Cas and -0.81 s/yr for V Lac.

Based on these two new O-C diagrams neither of them seems to show any association to a binary system.

**Résumé**

Nous donnons ici de nouveaux diagrammes d'O-C des deux céphéides classiques RY Cas et V Lac, établis sur une large base d'instants de maximum de lumière, qui couvre plus d'un siècle.

Ces deux céphéides présentent une variation de leur période pour laquelle il est possible de calculer le taux annuel. Ce taux vaut +1.86 s/an pour RY Cas et -0.81 s/an pour V Lac.

Sur la base de ces deux nouveaux diagrammes d'O-C l'appartenance à un système binaire est écarté pour ces 2 étoiles.

## Introduction

RY Cas and V Lac are two classical Cepheids more or less observed, although they have been discovered and observed since the end of the 19$^{th}$-century.

Table 1 gives the current GCVS elements for these two stars with the corresponding reference..

Table 1: GCVS elements.

| Star | RY Cas | V Lac |
|---|---|---|
| **HJD$_0$ (+2400000)** | 47419.26 | 28901.285 |
| **Period (days)** | 12.13856 | 4.983458 |
| **Coordinates J2000.0** | 23 52 07.04 +58 44 30.2 | 22 48 38.0 +56 19 17.6 |
| **Hipparcos number** | 117690 | 112626 |
| **HD number** | - | 240073 |
| **Reference of the ephemeris** | Bernikov 1992 | Erleksova 1982 |

Both have been identified as Cepheids with a strong variation of the period (Szabados 1977, 1991). On the basis of an update of the observations available in 2006, Meyer (2023a) gives some new accounts of the period rates for these two Cepheids (Table 3). We have given some new preliminary elements in our 2022 study (Boistel, 2022b), on the basis of new observations performed by automatized telescopes and visual observers since 2006.

This paper presents a major update for their O-C and a re-examination of their rate of period variation, as we did it in our previous study on the classical long-period Cepheid SV Vul (Boistel, 2022) for example.

1. **Observations and data compilations: times of maximum of light**

RY Cas and V Lac have been added to the visual program of the GEOS in order to complete the survey of the variation of their period. They have been selected among cepheids with strong period variations as established by Szabados (1977, 1991) in his extended surveys of northern Cepheids.

We have used a wide range of observations published in the classical literature of astrophysics and among CCD observations published by servers of automatized telescopes as ASAS-SN, Hipparcos catalogues, KWS, SWASP and typical database as the McMaster Cepheids database. We used also visual and CCD observations published by amateurs associations as AAVSO, BAV and GEOS (Boistel, 2022b) to complete our study.

Table 2 presents the time scale for the observations gathered and the number of times of maxima (ToM) collected. They represent twice the number of ToM collected by Meyer (2023a, 2023b).

Table 2: Time scale for the observations and number of observed ToM collected for this study.

| Star | RY Cas | V Lac |
|---|---|---|
| **JD range (+2400000)** | 16637-59777 | 05427-59741 |
| **Years** | 1904-2022 | 1873-2022 |
| **Number of days elapsed** | 43140 | 54315 |
| **Number of cycles covered** | 3553 | 10 899 |
| **Number of years covered** | 118 | 149 |
| **Number of times of maxima** | **87** | **145** |

Figure 1a shows the sources for the collected ToM for RY Cas, while figure 1b shows the sources explored for V Lac. The production of amateurs observers is quite important, and these stars have been well observed by the german association (BAV).

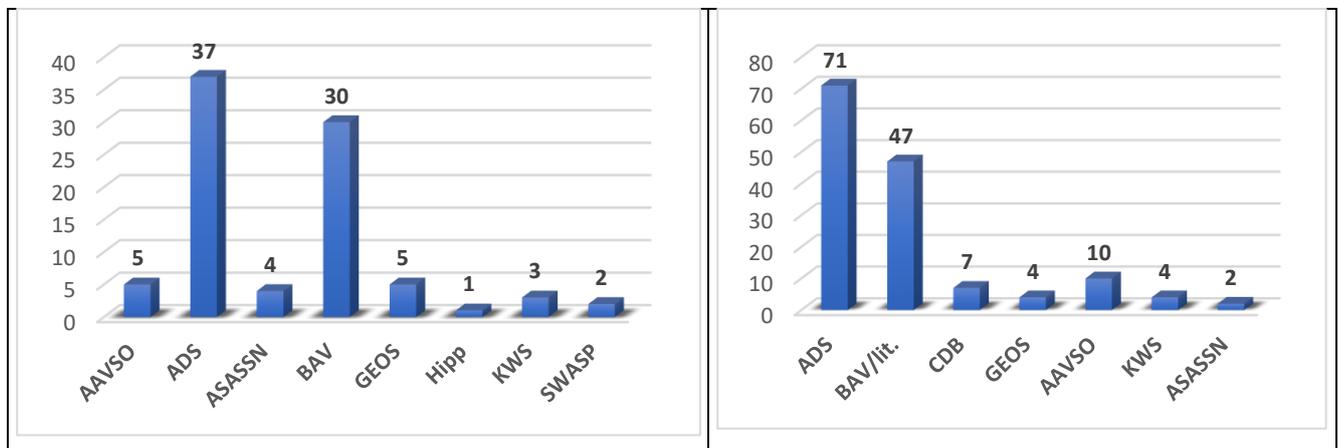

| Figure 1a: Sources for the 87 collected ToM of RY Cas. | Figure 1b: Sources for the 145 collected ToM of V Lac. |
|---|---|

Legend: ADS=Classic literature provided by NASA ADS service; ASASSN=All Sky Autom. Surv. for Super Novae data; AAVSO=American Association of Variable Stars Observers; BAV/Lit.=German Association of Variable Stars Observers (Bundesdeutsche Arbeitsgemeinschaft für Veränderliche Sterne) and BAV literature provided by ADS; CDB=McMaster Cepheids database; GEOS=Groupe Européen d'Observation Stellaire; Hipp=Hipparcos data; KWS= Kamogata/Kiso/Kyoto Wide-field Survey data; SWASP=Super Wide Angle Survey for Planets data).

## 2. Analysis: the O-C diagrams of RY Cas and V Lac and their interpretations

The light curves we can deduce from all the observations are regular and present a relative low dispersion (discrepancy). Figures 2a and 2b illustrate the average shape of the light curve of these two classical Cepheids. They are typical of their class of Cepheids (Meyer 2023b).

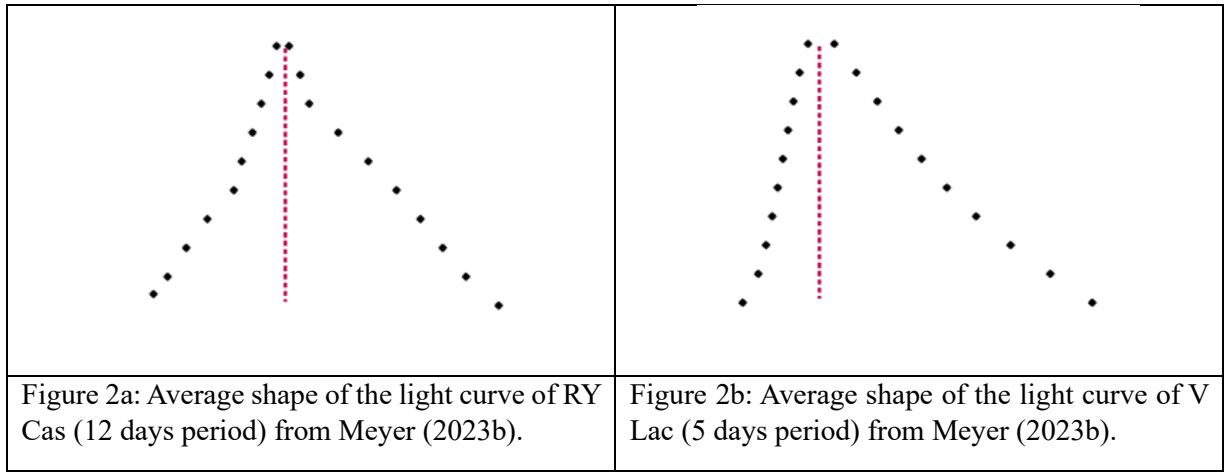

Figure 2a: Average shape of the light curve of RY Cas (12 days period) from Meyer (2023b).

Figure 2b: Average shape of the light curve of V Lac (5 days period) from Meyer (2023b).

The times of light maxima have been computed from the original collected observations with the help of two softwares Peranso© (Paunzen and Vanmunster, 2016) and MAVKA (Andrych and Andronov, 2020). These softwares provide various methods to fit the observations and to extract the times of light maximum.

The resulting O-C diagrams are shown on figures 3a and 3b. RY Cas has an increasing period while V Lac presents a decreasing period. Both present a quadratic variation of their period.

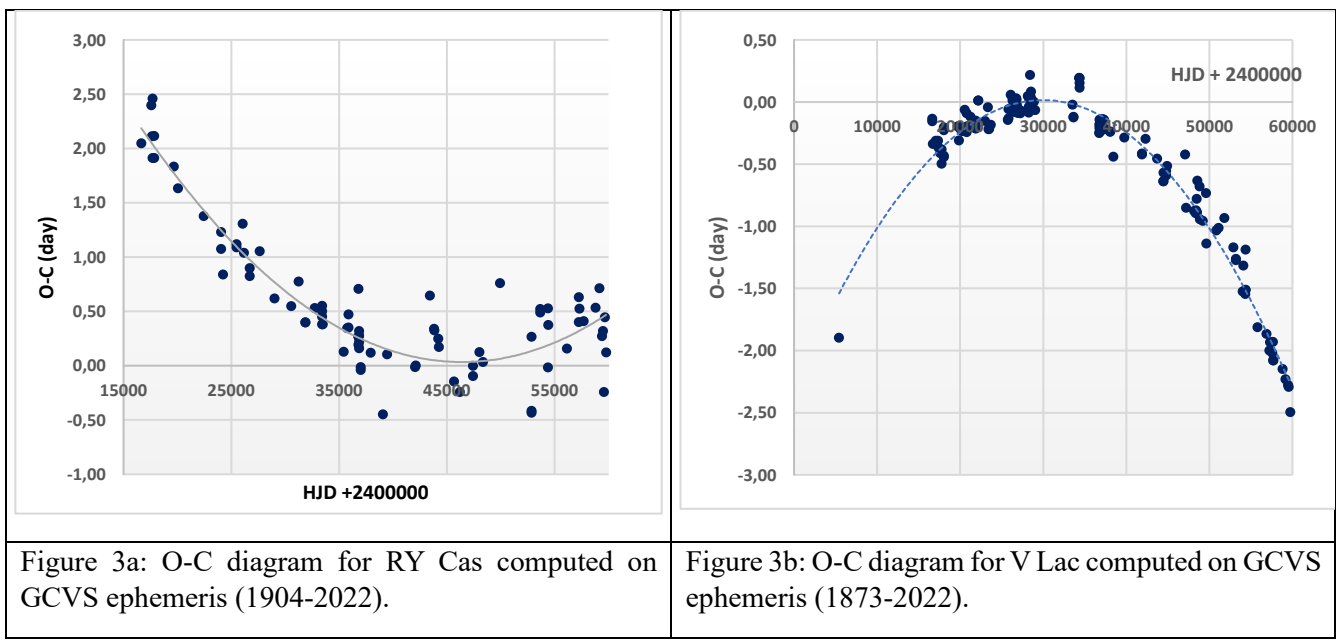

Figure 3a: O-C diagram for RY Cas computed on GCVS ephemeris (1904-2022).

Figure 3b: O-C diagram for V Lac computed on GCVS ephemeris (1873-2022).

For RY Cas, the mean of the quadratic residuals of the O-C is equal to -0.17 day while the standard deviation equals 0.592 day. For V Lac, these values are respectively equal to 0.011 and 0.117 day. We can see that the discrepancy is rather low for V Lac, a bit higher for RY Cas.

For RY Cas, five O-C residuals higher than 3 times the standard deviation, as usual, have been eliminated. They mainly come from early visual observations published by Luizet (Luizet, 1908).

For V Lac, two O-C residuals times of maxima higher than 3 times the standard deviation have been eliminated. The first comes from early observation by Blazhko given as a photographic time by Szabados (1977) but as a visual one by Meyer (2023). The second comes from the observatory of Wilno, is a visual estimate (Dziewulski is the observer) and given by Szabados (1977).

Nevertheless, we give these values in the list of the times of maxima on our download page for further studies.

The list of O-C are available to download on a GEOS-GitHub-Cepheids-deposit page at the following links: https://github.com/GEOS-Cepheids

**RY Cas files** (.txt, ;csv and .xlsx files): https://github.com/GEOS-Cepheids/RY-Cas

**V Lac files** (.txt, .csv and .xlsx files): https://github.com/GEOS-Cepheids/V-Lac

Some excerpts from these tables are put in appendices 1 and 2.

Table 3 presents the main results we can deduce from the analysis of all the O-C collected for each star: the new period with its standard error; the quadratic term computed with a quadratic fit; the period rate in s.yr$^{-1}$ computed with the formula explained below.

Table 3: Analysis of the data. Results of the O-C diagrams quadratic fittings.

| Star | RY Cas | V Lac |
|---|---|---|
| **Period P (days)** | 12.138678 ± 0.000045 | 4.9834857 ± 0.0000006 |
| **Log P** | 1.084171 | 0.697533 |
| **Quadratic term** | +3.82×10$^{-7}$ | -6.40×10$^{-8}$ |
| **Period rate (s.yr$^{-1}$) this paper (*)** | **+1.86** | **-0.81** |
| **Error on the period rate (log. diff. method) (s.yr$^{-1}$) this paper** | **±0.26** | **±0.04** |
| **Period rate (s.yr$^{-1}$) Meyer (2023a)** | +2.60 ±0.4 for 46 ToM | -0.82 ±0.04 for 77 ToM |

(*) The period rate has been computed with the following formula (Abdel-Sabour and Sanad, 2020):

$$\text{Period rate} = \frac{dP}{dt} [in\ s.yr^{-1}] = \left(\frac{2 \times Quad_{term}}{period}\right) \times 365.25 \times 24 \times 3600$$

Our new determinations of the period rates refine the results published by Meyer (2023a), in particular for **RY Cas**, considering the elimination of 5 undetermined times of maxima.

To explain these variations of period, let us summarized Evans' considerations (2015) as we did it in our GEOS Circular CEP4 (Boistel, 2022a) about the Cepheid RT Aur.

Evans et al. (2015) provides an account of the several causes suggested to explain period variations in Cepheids, both fundamental mode pulsators and first overtone pulsators:

(a) *Evolution through the Cepheids instability strip*: one direction of period change predominates and it results in a parabolic O-C diagram.
(b) *Light-time effects in binary systems:* this effect produces cyclic apparent period changes. But this must be consistent with the elements of the orbit of the system.
(c) *Star spots*: this cause has been suggested for the variations of the Kepler Cepheid V1154 Cyg. But while starspots could affect the time of maximum light, they would not have a cumulative effect as seen in the O-C diagram.
(d) *Mass-loss*: increasing periods and quadratic variations of the O-C observed in some Cepheids can be partially explained by mass-loss but it is only one of the several factors of the variations observed.
(e) *Pulsation and Blazhko effect*: if the phenomenon is not fully understood, pulsation and convection may drive pulsation mode excitation and hence amplitude modulation, and the same might also affect Cepheids periods.

For RY Cas and V Lac, we can't see any sign of additional periodic variation of the O-C diagrams shown on figures 3a and 3b.

### 3. Conclusion

From a systematic survey to collect new times of maxima for the stars RY Cas and V Lac, we can establish a period rate $\frac{dP}{dt}$ of (**+1.86 ± 0.26) s/yr for RY Cas** as a new result (increasing period) and **(– 0.81 ± 0.04) s/yr for V Lac** (decreasing period).
From the shape of the new O-C diagrams, none of these stars seems to belong to a binary system.

**Bibliography**

**Aknowledgements:**

The present study has made use of the following online tools and software facilities:
AAVSO VSTAR software: https://www.aavso.org/vstar-overview (Benn 2012)
AAVSO VSX The international variable stars index server: https://www.aavso.org/vsx/
ASAS-SN website: https://www.astronomy.ohio-state.edu/asassn/index.shtml
ADS bibliographical search: https://ui.adsabs.harvard.edu/classic-form
BAV data for scientists: https://www.bav-astro.eu/index.php/veroeffentlichungen/service-for-scientists/bav-data
McMaster Cepheid Photometry and Radial Velocity Data Archive:
https://physics.mcmaster.ca/Cepheid/.
KWS (Kamogata/Kiso/Kyoto Wide-field Survey): http://kws.cetus-net.org/
GCVS (Samus et al. 2017)
GEOS website: http://geos.upv.es/ and unpublished observations (private communications)
GEOS Open Access Publications: http://geos.upv.es/index.php/publications
MAVKA software: https://katerynaandrych.wixsite.com/mavka  (Andrych and Andronov 2020)
PERANSO software: https://www.cbabelgium.com/peranso/index.html (Paunzen & Vanmunster 2016)

**References (paper quoted and references of the published times of maxima to download):**

Abdel-Sabour, M., Sanad, M., 2020, NRIAG J. Astr. Geoph., 9, 1, 99
Andrych, K.D., Andronov, I. L., 2020, Journal of Physical Studies, vol. 24, no.1,1902
Benn, D., 2012, JAAVSO, 40, 852
Berdnikov, L.N., 1986, P.Z., vol. 22, 369
Berdnikov, L.N., 1987, P.Z., vol. 22, 530
Berdnikov, L.N., 1992a, Astron. and Astroph. Transac., vol.2, III, 1
Berdnikov, L.N., 1992b, Astron. and Astroph. Transac., vol.2, V, 43
Berdnikov, L.N., 1992c, Astron. and Astroph. Transac., vol.2, VI, 107
Berdnikov, L.N., 1992d, Astron. and Astroph. Transac., vol.2, VII, 157
Berdnikov, L. N., 1995, ASPC, 83, 349
Berdnikov, L.N., Mattei, J.A., Beck, S. J., 2003, JAAVSO, 31, 146
Boistel, G, 2022a, GEOS Circular on Cepheids, Cep 3 (SV Vul)
Boistel, G., 2022b, GEOS Circular on Cepheids, Cep 4 (RT Aur)
Boistel, G., 2022b, GEOS Note Circulaire NC 1318
Busquets, J., 1983, GEOS Note Circulaire NC 376
Efremov, I. N., 1978, Soviet Astronomy, vol. 22, 161

Eggen, O.J., 1951, Ap.J., 113, 367
Erleksova, G.E., & Irkaev, B.N., 1982, Peremennye Zvezdy, 21, 715
Evans N. R., et al., 2015, MNRAS, 446/4, 4008
Fadeyev, Yu. A., 2015, Astr. Let., 41/11, 640
Gaia Collaboration, 2018, VizieR Online Data Catalog.
Hübscher J., Lange Th., Paschke A., Vohla F., Walter F., 2005, OEJV, 1, 1
Hübscher J., Steinbach H.M., Th., Vohla F., Walter F., 2008, OEJV, 97, 1
Kiehl, M. & Hopp, U., 1977, IBVS 1329
Luizet, M., 1908, Bulletin astronomique, 25, 248
Madore, B.F. (ed.), 1985, *Cepheids: Theory and Observations*, Cambridge University Press.
Martin, C. & Plummer, H. C., 1916, MNRAS, vol. 76, 240
Meyer, R., 2023a, BAV Journal, 72
Meyer, R., 2023b, BAV Journal, 73
Meyer, R., 2023c, BAV Journal, 74
Milone, E.F., 1970, IBVS 482
Mitchell, R. I., Iriarte, B., Steinmetz, D., and Johnson, H. L., 1964, Boletin de los Observatorios Tonantzintla y Tacubaya, vol. 3, 153
Moffett T. J., Barnes T. G., 1984, ApJ Suppl., 55, 389
Oosterhoff, P. Th., 1960, Bull. Astron. Inst. Nether., 15, 501, 199
Parenago, P. P., 1957, «Communications of the Konkoly Observatory, 42, 53
Paunzen, E., Vanmunster, T., 2016, AN, 337, 239
Percy, J. R., 2021, JAAVSO, vol. 49, 1, 46
Pickering, E., 1914, Harvard Coll. Circ. 186, 1
Romano, G., 1955, Memorie della Societa Astronomica Italiana [MmSAI], vol. 26, 19
Samus, N.N., Kazarovets, E.V., Durlevich, et al., 2017, General Catalogue of Variable Stars: Version GCVS 5.1, Astronomy Reports, 61, No. 1, 80
Shapley, H., 1913, AN 194/4653, 20-21
Szabados, L., 1977, Communications of the Konkoly Observatory, 70, 1
Szabados, L., 1981, Comm. Konkoly Obs., 77, 1
Szabados, L., 1983, Astroph. Space Sci., 96/1, 185
Szabados, L., 1985, "Duplicity among the Cepheids in the northern hemisphere", in Madore, *op. cit.*, 75
Szabados, L., 1988, PASP, 100, 589
Szabados, L., 1991, Communications of the Konkoly Observatory, 96, 123
Szabados, L., 1996, A&A, 311, 189

# Appendices: excerpts from O-C lists to download

- Types of photometry legend:
  pe – photoelectric ; pg – photographic; V – Classic CCD Johnson V filter; vis. – visual ; un. – unknown type.
- Sources: See figures 1a and 1b legend.

**Appendice 1: O-C list for RY Cas, excerpt.**

| JJ Obs +2400000 | E GCVS | O-C (GCVS) | O-C AAVSO | O-C ADS | O-C BAV | O-C GEOS | O-C Hip | O-C KWS | O-C Swasp | Type Of Phot. | Source | Reference and observer (if known) |
|---|---|---|---|---|---|---|---|---|---|---|---|---|
| 16637,9200 | -2536 | 2,05 | | 2,05 | | | | | | pg | ADS | Szabados 1981 |
| 17560,8000 | -2460 | 2,40 | | 2,40 | | | | | | un. | BAV/Lit. | BAV Meyer 2023 |
| 17633,3460 | -2454 | 2,11 | | 2,11 | | | | | | vis | ADS | Szabados 1981 |
| 17658,6000 | -2452 | 3,09 | | 3,09 | | | | | | un. | BAV/Lit. | BAV Meyer 2023 |
| 17681,7000 | -2450 | 1,91 | | 1,91 | | | | | | un. | BAV/Lit. | BAV Meyer 2023 |
| Etc. | | | | | | | | | | | | |

**RY Cas - Link: https://github.com/GEOS-Cepheids/RY-Cas**

**Appendice 2: O-C list for V Lac, excerpt.**

| ToM HJD (+2400000) | E(GCVS) | O-C | O-C AAVSO | O-C ADS | O-C ASAS | O-C BAV | O-C CDB | O-C GEOS | O-C KWS | Type of Phot. | Source | Observer | reference |
|---|---|---|---|---|---|---|---|---|---|---|---|---|---|
| 5427,3000 | -4710 | -1,90 | | -1,90 | | | | | | un. | BAV/Lit. | un. | BAV Meyer 2023 |
| 16666,7400 | -2455 | -0,16 | | -0,16 | | | | | | pg | ADS | Oosterhoff | Oosterhoff, 1960 |
| 16666,7600 | -2455 | -0,14 | | -0,14 | | | | | | pg | ADS | Martin, Plummer | MNRAS jan 1916 |
| 16716,3920 | -2445 | -0,34 | | -0,34 | | | | | | vis | ADS | | Szabados 1977 |
| 17065,2590 | -2375 | -0,31 | | -0,31 | | | | | | vis | ADS | | Szabados 1977 |
| 17354,3000 | -2317 | -0,31 | | -0,31 | | | | | | un. | BAV/Lit. | un. | BAV Meyer 2023 |
| 17399,0970 | -2308 | -0,37 | | -0,37 | | | | | | vis | ADS | | Szabados 1977 |
| 17613,3410 | -2265 | -0,41 | | -0,41 | | | | | | vis | ADS | | Szabados 1977 |
| 17777,7100 | -2232 | -0,50 | | -0,50 | | | | | | un. | BAV/Lit. | un. | BAV Meyer 2023 |
| Etc. | | | | | | | | | | | | | |

**V Lac – Link: https://github.com/GEOS-Cepheids/V-Lac**